\documentclass[aps,groupedaddres8s,fleqn,nofootinbib,twocolumn]{revtex4-2}
\usepackage{graphicx}
\usepackage{xcolor}
\usepackage{amsmath,amssymb}
\usepackage{float}
\usepackage{amsfonts}
\usepackage{verbatim}
\usepackage{flushend}
\usepackage{balance}
\usepackage{endnotes}
\usepackage{footnote}
\usepackage{adjustbox}
\usepackage{gensymb}
\usepackage{float}
\usepackage{epigraph}
\usepackage{xcolor}
\usepackage[utf8]{inputenc}
\usepackage{setspace}
\usepackage[colorlinks=true,citecolor=blue,urlcolor=blue]{hyperref}

\newcommand{\beq}{\begin{eqnarray}}
\newcommand{\eeq}{\end{eqnarray}}
\usepackage{amsmath}

\begin{document}

\author{V. V. Brazhkin}
\affiliation{Vereshchagin Institute of High Pressure Physics, Russian Academy of Sciences, Kaluzhskoe shosse 14, Troitsk, Moscow, 108840, Russia}

\title{Density of states in liquids: quadratic or linear, and what each means?}

\begin{abstract}
A lot has been said about the vibrational density of states (DoS) in liquids. A more recent discussion introduces contradictions with earlier
results, and here I briefly review several pieces of evidence from modeling, experiments and theory showing this. I then show that the origin of
contradictions often comes from misinterpreting the nature of excitations in liquids. Classic quadratic DoS corresponds to propagating (albeit
damped) phonons as in solids and applies to any medium at low frequency, whereas the linear DoS comes from overdamped modes, localised particle
motions. This has implications for interpreting simulations, experiments and a theory of liquids. I also introduce a new mechanism where the
exponent of quadratic DoS reduces to close to 1 due to the asymmetry of the scattering intensity, reducing the effective phonon frequency.
\end{abstract}

\maketitle

1. The spectral density of states if often used to understand key properties of physical systems. It is understood in solids, gases, classical and
quantum. What is the vibrational density states in liquids? This is a particularly relevant question because the problem of theoretical desciption of
liquids was believed to be insurmountable due to the no-small parameter problem and hence unsolvable \cite{landaustat,Pitaevskii} (see Refs.
\cite{ropp} and \cite{mybook} for a review).

Liquids, just like solids, gases or many other systems, support sound - plane waves with the linear dispersion relation (DR) $\omega=ck$ enforced by
the wave equation \cite{landaustat}. From this dispersion relation, the quadratic vibrational density of states (DoS) generally follows without any
further assumptions \cite{landaustat}:

\begin{equation}
g(\omega)\propto\omega^2
\end{equation}

In liquids, the evidence for linear DR at low $\omega$ (as in solids) is numerous and comes from inelastic scattering experiments
\cite{copley,pilgrim,burkel,pilgrim2,water-tran,hoso,hoso3,monaco1,monaco2,sn,ruocco,ropp,mybook}. This was mostly done at low temperature where the
gap in $k$-space is small (see below). These experiments often concluded that propagating phonons in liquids and solids are remarkably similar in
terms of their DR. Accordingly, we used the Debye approximation extrapolating the quadratic DoS to the entire vibration spectrum as is done in the
solid theory \cite{landaustat} and calculated the phonon contribution to liquid energy and specific heat. This gave consistent understanding of the
experimental specific heat $c_v$ in a wide range of temperature and pressure \cite{ropp,mybook}. The resulting theory of liquid thermodynamics has
been thoroughly tested \cite{proctor1,proctor2}, used by many others to understand liquid thermodynamics and reviewed independently
\cite{chen-review,withbook}. Consistent with everything we know from experiments, theory and modelling of liquids, the universally seen decrease of
$c_v$ with temperature is due to the decrease of the number of propagating transverse phonons with temperature \cite{ropp,mybook} as discussed in
point 3 below.
\\

2. In view of wide generality of quadratic DoS, it is interesting that a {\it linear} DoS has been popping up in the liquid literature:

\begin{equation}
g(\omega)\propto\omega
\end{equation}

This discussion started in calculations using the instantaneous normal mode (INM) approach which takes an instantaneous liquid configuration,
diagonalizes dynamical matrix to obtain normal frequencies $\omega_i$ and forms their DoS. Some of $\omega_i$ are found to be real and some
imaginary. This typically gives linear DoS for real $\omega$ at high temperature $T$ and the crossover to the nonlinear quadratic DoS at low $T$
\cite{keyesli,otherquad} (other studies show the linear DoS at high temperature only).

At this point, I make two points about the INM method. First, the imaginary component of the normal frequency is unphysical (``inadmissible'' as
stated by Landau\&Lifshitz \cite{llmech}) because the imaginary component gives either exponentially increasing or decreasing terms for particle
displacements and hence non-conservation of energy. The INM approach seeks to get around this problem by considering short-time dynamics and explores
what can be learned \cite{keyes1}. Some issues are considered clarified whereas others aren't (eg including the the physical meaning of imaginary
modes, their interpretation and role in key physical effects) \cite{keyes1}. The second issue in the INM approach is that imaginary modes can be
present in the solid where this should not happen (see, e.g., Refs. \cite{path1,path3,path2}) and is therefore a pathology of the method. This can
simply come from a particle crossing the inflection point of the potential where the second derivative becomes negative (for the LJ potential, this
inflection point is located at the distance of about 1.1 of the equilibrium point only). This of course does not correspond to any instability and is
simply an artefact obfuscating the INM method and its results.

Of course there is another way to calculate the liquid spectra and vibrational frequencies which is free from the INM issues: this is based on
calculating current-current correlation functions \cite{boonyip,yangprl} which directly corresponds to what is measured in an inelastic scattering
experiment. These calculations give DRs that are non-surprisingly close to those measured in inelastic experiments, liner at low $\omega$ and hence
quadratic DoS as expected.

It is nevertheless interesting to ask why the INM approach gives the linear DoS at high $T$. Also interesting is the recent experiment reporting
linear DoS in liquids using the integration of the dynamical structure factor \cite{linearexp}, in seeming contradiction to the earlier experimental
evidence of linear DR and quadratic DoS. My discussion below is aimed at clarifying the INM, experimental
results and theory.\\

3. The first way to obtain the linear contribution to DoS comes from using DRs in liquids for longitudinal and transverse modes in liquids:

\begin{eqnarray}
\begin{split}
& \omega_l=c_lk\\
& \omega_t=\sqrt{c^2k^2-\frac{1}{\tau^2}}
\label{root}
\end{split}
\end{eqnarray}

\noindent where $l$ and $t$ subscripts refer to longitudinal and transverse waves, $\tau$ is liquid relaxation time. The square-root DR for
transverse waves and the associated gap in $k$-space are reviewed in Ref. \cite{gmsreview}. DoS corresponding to the second DR,
$g(\omega)\propto\frac{k^2}{{d\omega}/{dk}}$, and is \cite{ropp1}:

\begin{eqnarray}
\begin{split}
&g(\omega_t)\propto\omega^2\sqrt{1+\frac{1}{\omega^2\tau^2}}\\
&g(\omega_t)\propto\omega^2 ~{\rm for} ~\omega\tau\gg 1\\
&g(\omega_t)\propto\omega ~{\rm for} ~\omega\tau\ll 1
\label{cross}
\end{split}
\end{eqnarray}

Hence DoS is quadratic in the regime $\omega\tau\gg 1$. This is propagating phonon regime \cite{frenkel,ropp,mybook}. DoS is linear in the regime
$\omega\tau\ll 1$. This is the non-propagating regime where phonons are overdamped (their propagation range is much smaller than the wavelength)
\cite{frenkel,ropp,mybook} and where modes are localised on atoms and are not-oscillatory.

The linear DoS in the non-propagating regime enables us to interpret the nature of ``instantaneous modes'' in the INM calculation: modes with linear
DoS correspond to overdamped localised modes, whereas modes with quadratic DoS correspond to propagating phonons as expected in Eq. (1).

Hence, the modes for which $\omega\tau<1$ and $g(\omega)\propto\omega$ are overdamped nonpropagating nonvibrational modes and are localised on
individual atoms. Their contribution to liquid specific heat $c_v$ is trivial and purely kinetic as in the ideal gas state \cite{ropp,frenkel}. All
such ``modes'' only contribute to the total kinetic energy of the liquid $\frac{3}{2}Nk_{\rm B}T$ and to the kinetic part of liquid specific heat
$c_v=\frac{3}{2}k_{\rm B}$ \cite{ropp}. It is therefore unrelated to the universal experimental {\it decrease} of liquid $c_v$ with temperature
\cite{ropp} as might have been assumed \cite{bzpre}. This remains the case even if the linear DoS at low $\omega$ is modified to include dissipation
increasing with temperature \cite{bzpre} (see below).

Dissipation of course increases with temperature, however we know from experiments that it is not large enough to over-damp phonons at high
temperature. For example, inelastic scattering experiments show that the lifetime of the highest-frequency transverse waves in liquid Sn is about 1
ps, corresponding to $\omega>\Gamma=\frac{1}{\tau}$ ($\Gamma$ is decay/dissipation rate). The propagation range of these phonons is about 10 \AA\
comparable to the cage size and is comparable and in excess of the wavelength \cite{sn}. In liquid Fe, Cu and Zn, the lifetimes of these phonons are
in the similar sub-ps range for both transverse and longitudinal waves. This corresponds to the propagation range of about 9 \AA\ \cite{hoso3}. The
same is experimentally measured in liquid Ga, both at high and room pressure \cite{monaco2}. Very similar decay times and propagation ranges are
calculated in MD simulations \cite{yangprl}. In these experiments of which I only gave representative examples (see Ref. \cite{mybook} for a more
detailed review), the condition for the phonon to be propagating, $\omega\ge\Gamma$, applies. On the other hand, the criterion for the phonon to be
completely overdamped and localized, $\omega\ll\Gamma$, is not seen for these high-frequency phonons. Note that the above examples are the highest
frequency phonons where decay is the strongest. For lower-frequency, similar experiments show that the propagation condition $\omega\ge\Gamma$
applies with a much wider range of validity.

What's important to realise is that a very similar situation exists in {\it solids} where high-frequency phonons can also experience significant
dissipation and yet remain propagating and remain {\it not} overdamped. For example, the experimental lifetimes of 9 THz phonons in Si and Al are in
the range close to liquids: 0.3-0.4 ps in Al and 1 ps in Si at room temperature \cite{solids-decay}. Similar lifetimes are experimentally seen for 15
THz phonons in Si and GaN \cite{15thz1,15thz2}. In perovskites, the lifetime of high-frequency phonons is in the picosecond range \cite{perovpnas}
(see Ref. \cite{mybook} for review). The decay/damping of phonons in solids affects their transport properties, however it in no way changes the
fundamental premise of solid state physics that solid thermodynamic properties are due to phonons (with their decay coming on top as small anharmonic
corrections ) \cite{landaustat}. Each of these phonons in solids is decayed/damped to a degree but is {\it not} over-damped. As a result, the solid
specific heat is close to the Dulong-Petit value in the classical case - one of the central results in the solid state theory.

We saw that all experimental evidence shows, without exception, that the decay of propagating phonons in liquids is very close to that in solids.
This gives further support to the quadratic DoS in liquids (1) discussed earlier. These experimental results also negate previous assumptions
regarding the contribution of relaxing modes, rather than phonons, to liquid thermodynamics and liquid specific heat in particular
\cite{bzpnas,bzpre}.

There is only one essential difference between phonons in liquids and solids: {\it low-frequency} transverse phonons become overdamped below the
frequency $\frac{1}{\tau}$ as assumed by Frenkel \cite{frenkel} (the detailed underpinning theory, modelling and theory for this came in more
recently \cite{ropp,gmsreview,mybook}). These phonons are therefore removed from the liquid spectrum. Their energy becomes zero, and the kinetic
energy of jumping atoms emerges instead \cite{ropp,mybook}. The energy of remaining propagating high-frequency transverse modes can then be
calculated as an integral of the phonon energy, with DoS \eqref{cross} and the lower integration limit $\frac{1}{\tau}$ \cite{ropp,gmsreview,mybook}.
This lower limit means $\omega\tau>1$ for all frequencies to be integrated. This gives the quadratic DoS as is seen in Eq. \eqref{cross} which is
exactly what is used in the phonon theory of liquids \cite{ropp,mybook}.

The disappearance of transverse phonons with frequency below $\frac{1}{\tau}$ is then the key to the universal decrease of liquid $c_v$ with
temperature \cite{ropp,mybook}. However - and this is important point - this is the {\it only} effect that needs to be considered. The above
experiments show that transverse phonons which {\it remain propagating} in liquids above the frequency $\frac{1}{\tau}$ are indeed very close to
phonons in solids in terms of decay. At low temperature close to melting where $\frac{1}{\tau}$ in \eqref{root} is large and hence the number of
overdamped phonons is small, this picture predicts that $c_v$ should be close to $c_v$ in solids. The prediction is indeed what is widely seen
experimentally: liquid $c_v$ close to melting is
$c_v\approx 3k_{\rm B}$ \cite{wallacecv,ropp,mybook,proctor1,proctor2,nist}.\\

4. I now come back to the main point of this letter: DoS in liquids and its relation to what kind of excitations liquids really have. We saw that
those experiments returning the linear DR at low $\omega$ in liquids and hence quadratic DoS
\cite{copley,pilgrim,burkel,pilgrim2,water-tran,hoso,hoso3,monaco1,monaco2,sn,ruocco,ropp,mybook}
 are measuring propagating phonons. These are the ubiquitous collective excitations in liquids detected experimentally
(these excitations are to be contrasted to imaginary relaxing modes which were erroneously identified with the vibrational modes in Ref.
\cite{bzpnas} to derive the vibrational DoS in liquids).

The last point is welcome news: extensive previous and more recent experimental evidence above says that dynamics in liquids consists of two types of
motion: well-understood phonons and localised particle jumps, just like envisaged by Frenkel originally \cite{frenkel}. This means that liquids can
be consistently understood at the same level as solids are. There is no need for bringing in other notions such as imaginary or other types of modes
which may involve unclear physical concepts \cite{llmech}, can be prone to misinterpretations and involve incorrect conclusions \cite{bzpnas,bzpre}
(unless these notions are used, as a matter of convenience, to discuss trivial effects such as diffusive motion or localised atomic jumps
\cite{boonyip}).

Important to the above discussion is my explicit consideration of two key parameters: mode frequency $\omega$ and mode decay/dissipation parameter
$\Gamma=\frac{1}{\tau}$ and considering the parameter $\omega\tau$ setting the crossover between propagating phonon regime and overdamped localised
mode regime (it is this simple consideration which is often absent in the INM calculations). In a typical inelastic scattering experiment, decay of
measured excitations is always present and affects the dynamic structure factor $S(k,\omega)$. This decay is seen as the width of the intensity peak
as a function of $\omega$ at a given $k$. It therefore follows on general grounds that integrating $S(k,\omega)$ - as is done in the recent
experiment \cite{linearexp} - produces DoS which inherently contains the dissipation. 

It is important to note that at low $\omega$ (sub THz) not probed by inelastic scattering experiments (e.g., sound), dissipation of phonons is low
and can be ignored. Then, the quadratic DoS (1) follows readily from the linear DR \cite{landaustat} as discussed above. Therefore, extrapolating
this quadratic DoS to the entire range of frequencies in the phonon theory of liquids \cite{ropp} represents the same level of approximation as in
the Debye model
for solids \cite{landaustat}.\\

5. There is a specific mechanism related to decay/dissipation in liquids which moves the quadratic DoS close to linear which I now discuss.

In an inelastic scattering experiment (X-ray or neutron), a fixed $k$ corresponds to a set of $\omega$. If the scattered intensity is symmetric,
$\omega$ of an excitation at a given $k$ can be taken as a position of the peak of scattered intensity at that $k$. However, in liquids there is an
excess of low-frequency phonons as compared to high-frequency ones relative to the position of the frequency maximum (see, e.g. Figs. 8, 9, 11 in
Ref. \cite{ruocco}). If the intensity is integrated to get a weighted/average frequency, the weighted average is to the left of the frequency at the
peak and systematically reduces the weighted frequency compared to the frequency at the peak. A crude approximation of data describes this effect as
the reduction of the frequency exponent from $\omega\propto k$ in the symmetric case to about $\omega^{\frac{2}{3}}\propto k$ in the asymmetric case
(the detailed quantitative evaluation of this exponent is beyond the scope of this letter; my main aim is to discuss a mechanism based on the
asymmetry, whereas future work can provide more details and quantitative analysis.) This gives effective $\omega\propto k^{\frac{2}{3}}$ and the
linear {\it effective} DoS $g(\omega)\propto\frac{k^2}{{d\omega}/{dk}}\propto\omega$.

Of course this effective DoS corresponds to propagating phonons rather than localised excitations. What's important is that this spectra asymmetry
mechanism clearly interprets the nature of excitations in experimental work. By attributing these excitations to phonons rather than overdamped
localised modes (essentially local motions), it explains the relevance of propagating phonons for liquid physics including for thermodynamics,
transport and other phonon-related effects.

These and other issues related to liquid spectra will be discussed in my separate longer review.

I am grateful to K. Trachenko for discussing some key points in sections 1-4 above with me over the years, for bringing my attention to the INM
results, their interpretation and recent experiment in Ref. \cite{linearexp}.

\maketitle

%

\end{document}